\documentclass[11pt]{article}

\usepackage[dvips]{graphicx}
\usepackage{amsmath}
\usepackage{graphicx}
\usepackage{amsfonts}
\usepackage{amssymb}
\usepackage{epsfig}

\begin{document}

\title {Target space duality and moduli stabilization \\in String Gas Cosmology}
\author{Auttakit Chatrabhuti\\
\\
{\it Theoretical High-Energy Physics and Cosmology group}\\
{\it Department of Physics, Faculty of Science}\\
{\it Chulalongkorn University, Bangkok 10330, Thailand}\\
\\
{\it auttakit@sc.chula.ac.th} }
\date{\today}

\maketitle

\abstract{ Motivated by string gas cosmology, we investigate the
stability of moduli fields coming from compactifications of string
gas on torus with background flux.  It was previously claimed that
moduli are stabilized only at a single fixed point in moduli
space, a self-dual point of T-duality with vanishing flux.  Here,
we show that there exist other stable fixed points on moduli space
with non-vanishing flux.  We also discuss the more general target
space dualities associated with these fixed points.
}

\section{Introduction}

String gas cosmology pioneered by Brandenberger and Vafa \cite{BV}
is one of the attempts to apply string theory in cosmology. The
advantage of this model is that it can provide a solution to the
initial singularity problem and can explain the dimensionality of
space-time.  The universe in string gas cosmology starts from a
very small, dense, and hot state where the matter content is
dominated by gas of closed strings. All 9-dimensional spatial
dimensions are taken to be toroidal compactified at the radius of
string length.

The target space duality (T-duality) \cite{T-dual} in string
theory implies the minimum length scale that we can probed. This
give us a possible solution to avoid the initial singularity at
the time of the big bang.  The importance of the winding modes
corresponding to closed strings wrapped around a circle of
compactified dimension is also recognized.   To get the model with
expanding universe, the winding modes need to be annihilated.  The
dimensionality of the universe is the result of decompactification
of 3 out of 9 spatial dimensions through winding modes
annihilation process \cite{TV,KP}. The extension of this
cosmological consideration to toroidal orbifolds was considered in
\cite{Greene1}. These nice features make string gas very
attractive.  Its brane generalization is investigated in
\cite{alex, WB1, Boehm, Easson, Bassett, Campos, Greene2, Trodden,
Kaya}.

The purpose of this paper is to investigate the stability of the
the moduli fields which describe the volume and the shape of the
unobserved compact extra-dimensions and the background
antisymmetric field.  This is known as the moduli problem in
string gas cosmology (for a review see \cite{moduli}).  Previous
works on moduli stabilization of string gas compactification show
that the moduli can be stabilized at the self-dual fixed-point
with vanishing antisymmetric field.  Numerical evidence for
stability of the volume moduli was shown by Brandenberger and
Watson but they find that the dilaton is running logarithmically
\cite{WB2}. The effective field theory approach shows that the
dilaton and the radion can not be stabilized except for the
5-dimensional case \cite{Battefeld}. The important roles of string
massless modes in stabilization mechanism was discovered by Patil
and Brandenberger \cite{Patil} and the stability of the volume
moduli can be proved analytically.  The Higgs-like mechanism for
stabilizing moduli at points of enhanced gauge symmetry was
proposed in \cite{Watson}. Recently, Kanno and Soda \cite{Soda}
considered the 4-dimentional effective action by taking into
account T-duality. They have shown that the dilaton is marginally
stable as the dilaton potential disappears when the moduli is
stabilized at the self-dual radius.  In this paper, we suggest
there should exist other stable fixed-points on the moduli space
of string gas compactification with non-zero anti-symmetric field
background.

This paper is organised as follows: in section \ref{sec:lattice }
we review the compactification of closed and heterotic string on a
torus with backgroud antisymmetric field by following the works of
Narain et al. \cite{Narain, NSW}.  In section \ref{sec:fixed
point} we discuss target space duality symmetries of toroidal
compactification and their fixed-points. We express the 4
dimensional T-duality invariant effective action in section
\ref{sec:effective}. In section \ref{sec:stability} by considering
4 dimensional effective action for a gas of heterotic string
compactified on $T^{2}\times T^{2}\times T^{2}$, we show the
stability of moduli at the fixed-points. We also discuss the
stability of the dilaton field. In section \ref{sec:conclusions}
we present our conclusions.

\section{Closed string on lattice torus }
\label{sec:lattice }

Consider a closed string propagating on a torus with metric
$G_{IJ}$ and a background anti-symmetric field $B_{IJ}$ described
by the action:

%
\begin{equation}
 S = \frac{1}{4\pi\alpha'}\int d\tau d\sigma \left[\sqrt{-\gamma}\gamma^{ab}G_{IJ}\partial_{a}X^{I}\partial_{b}X^{J}
 +\epsilon^{ab}B_{IJ}\partial_{a}X^{I}\partial_{b}X^{J} \right],
\label{world-sheet}
\end{equation}
where $X^{I}$ ($I=1\cdots d$) represent the toroidal coordinates
and we ignore the non-compact directions for the moment.  As we
will see later, it is convenient to encode all the geometric data
of the compact background in the "background matrix" $E$ where
\begin{equation}
E_{IJ} = G_{IJ}+B_{IJ}. \label{E_matrix_def}
\end{equation}
Let us take the world-sheet metric $\gamma^{ab}=\eta^{ab}$ and
$X^I$ to be normalized such that $X^I$ and $X^I+2\pi n^I$ are the
same point on the torus for any integer winding number $n^I$. The
canonical momentum $P_I$ associated with $X^{I}$ is given by
\begin{equation}
2\pi\alpha'P_{I} = G_{IJ}\dot{X}^{J} + B_{IJ}X'^{J} = p_{I} +
oscillators, \label{conjugate_momentum}
\end{equation}
where the string center of mass momentum $p_I$ is quantized in
integer units $p_I = m_I$ due to the periodicities of $X^{I}$
coordinates.  $m_I$ is the momentum number.  The coordinates
$X^{I}$ can be split into left-moving part and right- moving part,
$X^{I}= X_{L}^{I} + X_{R}^{I}$. By following \cite{NSW} the mode
expansions for $X_L^I$, and $X_R^I$ can be written as

\begin{eqnarray}
X_L^I &=& \frac{x_L^I}{2} +
\sqrt{\frac{\alpha'}{2}}\alpha_{0}^{I}(\tau+\sigma)+
i\sqrt{\frac{\alpha'}{2}}\sum_{n\neq
0}\frac{1}{n}\alpha_n^Ie^{-in(\tau+\sigma)},\\
X_R^I &=& \frac{x_R^I}{2} +
\sqrt{\frac{\alpha'}{2}}\tilde{\alpha}_{0}^{I}(\tau-\sigma)+
i\sqrt{\frac{\alpha'}{2}}\sum_{n\neq
0}\frac{1}{n}\tilde{\alpha}_n^Ie^{-in(\tau-\sigma)}.
 \label{mode_expand}
\end{eqnarray}
The background fields $G_{IJ}$ and $B_{IJ}$ only affect the
quantization of string zero-modes which can be expressed in terms
of left and right moving momenta,
 $\alpha_0^I=\sqrt{\frac{\alpha'}{2}}p_L^I$ and
$\tilde{\alpha_0}^I=\sqrt{\frac{\alpha'}{2}}p_R^I$, with
\begin{eqnarray}
p_L^I &=& G^{IJ}m_{J}+\frac{n^{I}}{\alpha'}-G^{IJ}B_{JK}\frac{n^{K}}{\alpha'},\nonumber\\
p_R^I &=&
G^{IJ}m_{J}-\frac{n^{I}}{\alpha'}-G^{IJ}B_{JK}\frac{n^{K}}{\alpha'},
\label{zero_mode}
\end{eqnarray}
The mode expansion of conjugate momenta reads
\begin{equation}
2\pi\alpha'P_{I} = p_{I} + \sqrt{\frac{\alpha'}{2}}\sum_{n\neq
0}[E_{IJ}\alpha_{n}^{J}e^{-in(\tau+\sigma)} +
E_{IJ}^{T}\tilde{\alpha}_n^Je^{-in(\tau-\sigma)} ].
\label{expand_momentum}
\end{equation}
By expanding the equal-time commutation relation
$[X^{I}(\sigma,0),P_{J}(\sigma',0)]=i\delta^{I}_{J}\delta(\sigma-\sigma')$,
the non-zero commutation relations for $\alpha_{n}^{I}$,
$\tilde{\alpha}_n^I$, $x^{I} = \frac{1}{2}(x^{I}_{L}+x^{I}_{R})$
and $p_{I}$ are
\begin{equation}
[x^{I},p_{J}]=i\delta^{I}_{J}\;\;\;,\;\;\;
[\alpha_{n}^{I},\alpha_{m}^{J}]=[\tilde{\alpha}_n^I,\tilde{\alpha}_m^J]
= mG^{IJ}\delta_{m+n,0}. \label{commutation}
\end{equation}
Here the oscillators and their commutation relation are background
dependent.  The Virasoro generators are defined by
\begin{equation}
L_{0}= \frac{1}{2}\alpha_{0}^{I}\alpha_{0}^{I} + N\;\;\;,\;\;\;
\tilde{L}_{0} = \frac{1}{2}\tilde{\alpha}_0^I\tilde{\alpha}_0^I +
\tilde{N}, \label{Virasoro}
\end{equation}
where the number operators:
\begin{equation}
N = \sum_{n > 0}\alpha_{-n}^{I}G_{IJ}\alpha_{n}^{J}\;\;\;,\;\;\;
\tilde{N} = \sum_{n >
0}\tilde{\alpha}_{-n}^IG_{IJ}\tilde{\alpha}_n^J.
\label{number_operator}
\end{equation}

we can define vectors $e^{i}_I$ to be a basis to the
compactification lattice $\Lambda^{d}$, such that the
$d$-dimensional torus is $T^{d}=R^{d}/\pi\Lambda^{d}$.  A basis to
the dual lattice $\Lambda^{d*}$ is denoted by $e^{*I}_{i}$.  The
indices $i,j$ label an orthonormal basis to the torus.  The scalar
products of vectors $e^{i}_I$ and $e^{*I}_{i}$ are:

\begin{equation}
\sum_{i = 1}^{d}e_{I}^{i}e_{J}^{i} = G_{IJ},\;\; \sum_{i =
1}^{d}e_{i}^{*i}e_{i}^{*J} = G^{IJ},\;\; \sum_{i =
1}^{d}e_{I}^{i}e_{i}^{*J} = \delta_{I}^{J}.\label{basis}
\end{equation}
In this basis, the momenta in (\ref{zero_mode}) are:
\begin{eqnarray}
p_{Li} &=& [m_{I}+ E^{T}_{IJ}\frac{n^{J}}{\alpha'}]e^{*I}_{i},\nonumber\\
p_{Ri} &=& [m_{I}- E_{IJ}\frac{n^{J}}{\alpha'}]e^{*I}_{i},
\label{m_lattice}
\end{eqnarray}
To ensure modular invariance of the closed string spectrum, it is
required that $\alpha_{0}^{2} - \tilde{\alpha}_{0}^{2} =
m^{I}n_{I}\in 2 \mathbb{Z}$. This implies that the vectors
$(\sqrt{\frac{\alpha'}{2}}p_{Li},\sqrt{\frac{\alpha'}{2}}p_{Rj})$
span even self-dual $(d,d)$ Lorentzian lattice $\Gamma^{(d,d)}$,
with negative (positive) signature for left (right) momenta. The
mass formula for closed string is given by:
\begin{equation}
m^{2}=\frac{1}{2}(p_{L}^{2} +
p_{R}^{2})+\frac{2}{\alpha'}(N+\tilde{N}-2), \label{closed_mass}
\end{equation}
with the level matching condition $L_{0}-\tilde{L}_{0} = 0$ which
can be written as:
\begin{equation}
N-\tilde{N} = m_{I}n^{I}. \label{closed_matching}
\end{equation}
For generic toroidal compactifications, the massless vectors
$\alpha_{-1}^{\mu}\tilde{\alpha}_{-1}^{I}\left|0_{L},0_{R}\right>$
and
$\tilde{\alpha}_{-1}^{\mu}\alpha_{-1}^{I}\left|0_{L},0_{R}\right>$,
where the $\mu$ and $I$ indices respectively label the non-compact
and compact dimensions, generate a local
$[U(1)]^{d}\times[U(1)]^{d}$ gauge symmetry.  As it was first
pointed out in \cite{Narain}, more massless vectors arise when
$\sqrt{\frac{\alpha'}{2}}p_{Li}$ and
$\sqrt{\frac{\alpha'}{2}}p_{Ri}$ are roots of the simply laced
group $\mathcal{G}_L$ and $\mathcal{G}_R$ of rank $d$ with root
length $\sqrt{2}$.  The gauge symmetry is enhanced to
$\mathcal{G}_L\times \mathcal{G}_R$.

Note that in the case of heterotic string, we can include 16
left-moving bosons propagating in maximal torus of $E_8\times E_8$
or $Spin(32)/\mathbb{Z}_{2}$. The resulting $\Gamma^{(16+d,d)}$
lattice is still an even self-dual Lorentizian lattice.  However,
this extra degree of freedoms will not affect our later
considerations. As you will see later in section
\ref{sec:stability}, it is enough for us to concentrate on the
$d\times d$ dimensional sublattice of the full $
\Gamma^{(16+d,d)}$ lattice. The mass formula for heterotic string
is:
\begin{equation}
m^{2}=\frac{1}{2}(p_{L}^{2}+p_{R}^{2})+\frac{2}{\alpha'}(N+\tilde{N}-1),
\label{hmass_formula}
\end{equation}
with matching condition
\begin{equation}
N-\tilde{N} + 1 = m_{I}n^{I}. \label{h_matching}
\end{equation}
Note that in this case $N$ and $\tilde{N}$ are the number
operators for (26-dimensional) left-handed oscillators and
(10-dimensional) right-handed oscillators respectively.  We can
also consider enhancement of gauge symmetry from heterotic string
compactified on $d$-dimensional torus. For generic points in
moduli space, the gauge symmetry is $U(1)_{L}^{16+d}\times
U(1)_{R}^{d}$.  However some of the $U(1)$ can be enhanced to
non-abelian gauge groups.  This will happen when the lattice
$\Gamma^{(16+d,d)}$ contains vectors with
$(\sqrt{\frac{\alpha'}{2}}p_{L})^{2}=2$ and
$\sqrt{\frac{\alpha'}{2}}p_{R}=0$.  The vectors
$\sqrt{\frac{\alpha'}{2}}p_{Li}$ are roots of the simply laced
group $\mathcal{G}_L$ of rank $16+d$ with root length $\sqrt{2}$.
The largest enhanced symmetry gauge group that we can have is
$\mathcal{G}_L = SO(32+2d)$.

\section{Target space duality and their fixed points}
\label{sec:fixed point}

We are now at the position to discuss the target space duality in
closed string theory and examine their fixed-points.  The target
space duality will work in a similar way as in the heterotic
string case.  Let us start by rewriting closed string mass formula
(\ref{closed_mass}) in a $O(d,d,\mathbb{Z})$ covariant form (in
$\alpha'=1$ unit) as:
\begin{equation}
m^{2}= Z^{T}M(E)Z + 2(N + \tilde{N} - 2). \label{odd_mass}
\end{equation}
The matching condition is also rewritten as $N - \tilde{N} =
\frac{1}{2}Z^{T}\eta Z$, where we define
\begin{equation}
M(E) = \left(%
\begin{array}{cc}
  G^{-1} & -G^{-1}B \\
  BG^{-1} & G-BG^{-1}B \\
\end{array}%
\right)\;\;,\;\; \eta = \left(%
\begin{array}{ll}
  0 & \mathbf{1}_{d\times d} \\
  \mathbf{1}_{d\times d} & 0 \\
\end{array}%
\right),  \label{M_matrix}
\end{equation}
and $Z = (m_{1},\ldots, m_{d}, n_{1},\ldots, n_{d})$.  It is
obvious that the closed string spectrum is invariant under
$O(d,d,\mathbb{Z})$ transformation:
\begin{equation}
M \rightarrow \Omega M \Omega^{T}\;\;,\;\; Z \rightarrow \Omega Z,
\label{odd_trans}
\end{equation}
where $\Omega \in O(d,d,\mathbb{Z})$ is the integer valued matrix
satisfying $\Omega^{T}\eta\Omega = \eta$.

We would like to discuss a particular element $\Omega = \eta$ of
$O(d,d,\mathbb{Z})$.  This is known as the T-duality. It is given
by the inversion of the background matrix $E$ \cite{GRV,SW}:
\begin{equation}
E \rightarrow E' = G'+B'= E^{-1}.
\label{E_duality_def}
\end{equation}
Here we take $G'$ and $B'$ to be the symmetric and antisymmetric
part of $E'$ respectively.  Note that the duality maintains the
symmetry (anti-symmetry) of $G$ ($B$). Under T-duality the metric
and the antisymmetric field transform as
\begin{eqnarray}
G \rightarrow G' = (G-BG^{-1}B)^{-1}, B \rightarrow B' = (B-GB^{-1}G)^{-1},\nonumber\\
G^{-1}B \rightarrow -BG^{-1}. \label{E_duality_m}
\end{eqnarray}
The closed string mass spectrum (\ref{closed_mass}) is manifestly
invariant under transformation (\ref{E_duality_m}) together with
the interchange of winding modes with momentum modes.

Let us consider the fixed point of T-duality by starting with the
special case where the antisymmetric $B$ field is absence.  We
obtain the transformation:
\begin{equation}
G \rightarrow G^{-1}. \label{G_inverse}
\end{equation}
For one compact dimension ($d = 1$), one can think of a closed
string moving in the circle of radius $R$. T-duality $R
\rightarrow \frac{1}{R}$ has a fixed point at the self-dual radius
$R = 1$.  At this fixed point the gauge symmetry is enlarged form
$U(1)_{L}\times U(1)_{R}$ to $SU(2)_{L}\times SU(2)_{R}$.

For more general compactification, the $E\rightarrow E^{-1}$
duality still has a single fixed point at
\begin{equation}
G = \mathbf{1}_{d\times d},\;\;\;B = 0, \label{G_fixed_point}
\end{equation}
which is the unique solution to the equation
$(G+B)^{2}=\mathbf{1}_{d\times d}$ for positive definite $G$. We
can show that the gauge symmetry get enhanced to $SU(2)_L^d\times
SU(2)_R^d$.

However, there exists a more general example of fixed-points of
T-duality modulo $SL(d,\mathbb{Z})$ and $\Theta(\mathbb{Z})$
transformation \cite{GPR, GRV, SW}:
\begin{equation}
E^{-1} = M^{T}(E + \Theta)M,\:,\; M \in SL(d,\mathbb{Z}),\; \Theta
\in \Theta(\mathbb{Z}),\label{E_mod}
\end{equation}
where $\Theta(\mathbb{Z})$ is the symmetry transformation that
adds to $B_{IJ}$ an antisymmetric integral matrix $\Theta_{IJ}$
($\Theta_{IJ} \in \mathbb{Z}$ and $\Theta_{IJ} = - \Theta_{JI}$ ).
A class of background with a maximally enhanced symmetry gives an
example of such fixed points \cite{EGRS}. This background are
characterized by $e^{i}_{I} = \sqrt{\frac{\alpha'}{2}}r^{i}_{I}$
where the vector $r^{i}_{I}$ span the root lattice
$\Lambda_{root}$ of $\mathcal{G}$ and are chosen to be the simple
roots with $(r_{I})^{2}=2$.  It follows from (\ref{basis}) that
$\frac{2}{\alpha'}G_{IJ}$ is the Cartan matrix of $\mathcal{G}$.
To be more precise, we choose the metric and the antisymmetric
field as:

\begin{equation}
G_{IJ} = \frac{1}{2}C_{IJ},\;\;
B_{IJ} = \left\{%
\begin{array}{cl}
  G_{IJ} & , I<J \\
  0 & , I=J \\
  -G_{IJ} & , I>J \\
\end{array}%
\right., \label{E_fixed_point}
\end{equation}
where $C_{IJ}$ is the Cartan matrix of $\mathcal{G}$.  Since $E,
E^{-1} \in SL(d,\mathbb{Z})$, the solution to (\ref{E_mod}) is
obtained by taking $M = E^{-1}$ and $\Theta = E^{T}-E$.

\begin{figure}[!h] \centering
  \includegraphics[width=0.5\textwidth, bb = 0 0 420 332 ]{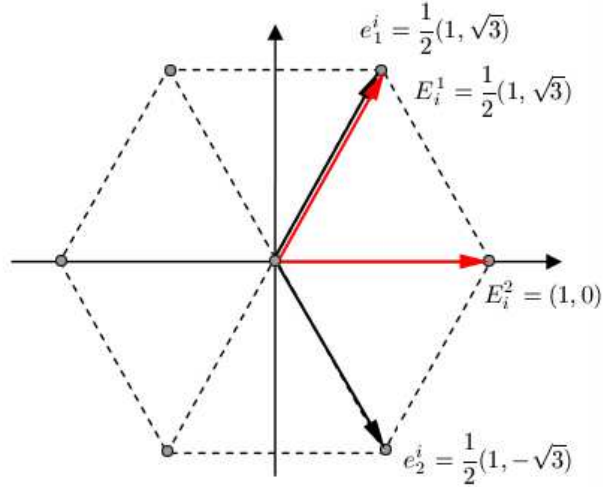}\\
  \caption{The basis $e_I^i$ and the E-dual basis $E_i^I$ span the same $SU(3)$ lattice.}
  \label{fig:diagram}
\end{figure}

Let us consider the $d = 2$ example. There are two fixed points of
enhanced symmetry.  The first is the fixed point of
$SU(2)_L^{2}\times SU(2)_R^{2}$ symmetry as in
(\ref{G_fixed_point}).  The second is the $SU(3)_{L}\times
SU(3)_{R}$ fixed point which the background matrix $E$, $G$ and
antisymmetric field $B$ are:

\begin{equation}
E_{IJ} = \left(%
\begin{array}{cc}
  1 & -1 \\
  0 & 1 \\
\end{array}%
\right),\;\;G_{IJ} = \frac{1}{2}\left(%
\begin{array}{cc}
  2 & -1 \\
  -1 & 2 \\
\end{array}%
\right),\;\; B_{IJ} = \frac{1}{2}\left(%
\begin{array}{cc}
  0 & -1 \\
  1 & 0 \\
\end{array}%
\right), \label{sample_fixed_point}
\end{equation}
or equivalently, we can choose the basis vector in (\ref{basis})
to be:
\begin{equation}
e_{1}^{i} = \frac{1}{2}(1, \sqrt{3}),\; \; e_{2}^{i} =
\frac{1}{2}(1, -\sqrt{3}).\label{su3_basis}
\end{equation}
Under duality (\ref{E_duality_def}), we have
\begin{equation}
E^{-1}_{IJ} = \left(%
\begin{array}{cc}
  1 & 1 \\
  0 & 1 \\
\end{array}%
\right),\;\;G'_{IJ} = \frac{1}{2}\left(%
\begin{array}{cc}
  2 & 1 \\
  1 & 2 \\
\end{array}%
\right),\;\; B'_{IJ} = \frac{1}{2}\left(%
\begin{array}{cc}
  0 & 1 \\
  -1 & 0 \\
\end{array}%
\right). \label{su3_dual}
\end{equation}
It is obvious that $E^{-1}\neq E$ at this fixed point.  The
importance of this fixed point can be observed by considering the
effect of T-duality on the $SU(3)$ basis vectors.  This effect is
called the E-duality transformation in \cite{auttakit} and the
basis vectors are transformed as:
\begin{equation}
e_{I}^{i} \rightarrow E_{i}^{I},\; \; e_{i}^{*I}\rightarrow
E_{I}^{*i}. \label{basis_change}
\end{equation}
Here we can choose:
\begin{equation}
E_{i}^{1} = \frac{1}{2}(1,\sqrt{3}),\; \; E_{i}^{2}=(1,0).
\label{dual_su3}
\end{equation}
One can see that the basis (\ref{su3_basis}) and the E-dual basis
(\ref{dual_su3}) span the same $SU(3)$ lattice (see
Figure\ref{fig:diagram}).  Hence, T-duality transformation changes
only the basis vectors of the lattice without changing the lattice
itself.

In the case of heterotic string compactified on $d=2$ dimensioanl
torus, the background in (\ref{G_fixed_point}) is associated with
the self-dual fixed point with enhanced gauge symmetry
$SU(2)\times SU(2)$.  The metric and antisymmetric field in
(\ref{sample_fixed_point}) correspond to the fixed-point with
$SU(3)$ enhanced gauge symmetry.  We will call them them the
$SU(2)\times SU(2)$ and the $SU(3)$ fixed-point respectively.

\section{Low energy effective action for string gas}
\label{sec:effective}

The bosonic part of the low energy effective action for closed
string is given by
\begin{equation}
S = -\frac{1}{2\kappa^{2}}\int d^{10}x
\sqrt{-G_{10}}\;e^{-2\phi}(R^{(D)}+4\partial^{A}
\phi\partial_{A}\phi-\frac{1}{12}H^{2}), \label{eff_action_1}
\end{equation}
where $H=dB$ is the field strength for the antisymmetric field,
$\kappa$ is the 10 dimensional gravitational coupling constant.
$G_{10}$ and $\phi$ represent the 10-dimensional metric and the
dilaton respectively. The indices $A = 0,1,\ldots,9$.

There are some assumptions we would like to make. First, we apply
the Brandenberger-Vafa mechanism such that the four dimensional
space-time is practically non-compact while the 6-dimensional
internal space is toroidally compactified. Second, we assume the
cosmological ansatz :
\begin{equation}
ds^{2} =
g_{\mu\nu}(x^{\mu})dx^{\mu}dx^{\nu}+G_{IJ}(x^{\mu})dy^{I}dy^{J} .
\label{cosmological_metric}
\end{equation}
Here, the metric $G_{IJ}$ represent the d-dimensional compact
space and $g_{\mu\nu}$ is the metric for the 4 dimensional
non-compact space-time with $\mu,\nu = 0,\ldots,4$.  We also
assume the metric depends only on the non-compact coordinates
$x^{\mu}$. For simplicity, we choose the anti-symmetric field to
have non-vanishing components only in the compact directions.  The
action (\ref{eff_action_1}) can be written as:
\begin{equation}
S = -\frac{V_{6}}{2\kappa^{2}}\int d^{4}x
e^{-\Phi}(R^{(4)}+\partial_{\mu}\Phi\partial^{\mu}\Phi +
\frac{1}{8}Tr[\partial_{\mu}M\eta\partial^{\mu}M\eta]) ,
\label{eff_action_2}
\end{equation}
where the shift dilaton filed is defined by:
\begin{equation}
\Phi = 2\phi - \ln\sqrt{G} . \label{shift_dilaton}
\end{equation}
It is easy to show that the action (\ref{eff_action_2}) is
invariant under T-duality \cite{GV}.

The action for a gas of string is given by
\begin{equation}
 S_{gas} = \mu_{4}\int d^{4}x \sqrt{-g_{00}}\;\rho,
\label{gas_action}
\end{equation}
where $\mu_{4}$ is the comoving number density of a string gas and
the energy $\rho$ is defined by
\begin{equation}
 \rho = \sqrt{g^{\mu\nu}p_{\mu}p_{\nu} + m^{2}(E)}.
\label{gas_energy}
\end{equation}
It is easy to see that the string gas action is also T-duality
invariant. In order to consider the stability, we choose to work
in the Einstein frame by defining the Einstein metric
$(g_{E})_{AB} = e^{-\Phi}g_{AB}$. The full action in the Einstein
frame is
\begin{eqnarray}
 S_{E} = &-&\frac{V_{6}}{2\kappa^{2}}\int d^{4}x\sqrt{-g_E}\;
(R^{(4)}_{E}+g_E^{\mu\nu}\partial_{\mu}\Phi\partial_{\nu}\Phi +
\frac{1}{8}Tr[\partial_{\mu}M\eta\partial^{\mu}M\eta]) \nonumber\\
&+& \int d^{4}x \sqrt{-g_{E}}\;V_{eff}(g_{E},\Phi,E),
\label{total_action}
\end{eqnarray}
where the effective potential is,
\begin{equation}
 V_{eff}(g_{E},\Phi,E) \equiv \frac{\mu_{4}}{\sqrt{g_{E}^{s}}}\sqrt{g_{E}^{\mu\nu}p_{\mu}p_{\nu} +
e^{\Phi}m^{2}(E)}. \label{effective_potential}
\end{equation}

\section{Stability of heterotic string gas compactified on torus}
\label{sec:stability}

In order to avoid instability from the presence of tachyonic
states \cite{Witten}, we study a model of supersymmetric string
gas. Since toroidal compactification of Type II superstring does
not give us enhanced gauge symmetry, we will consider the SO(32)
heterotic string gas compactified on $T^{2}\times T^{2}\times
T^{2}$.

Concentrating on the compact part, we can analyze each torus
separately.  As we have seen in the $d=2$ example, there must
exist two fixed points with enhanced symmetry in the moduli space
of each 2-dimensional torus. The first is the $SU(2)\times SU(2)$
fixed-point and the second is the $SU(3)$ fixed point.  We would
like to show that these two fixed points are stable.

We will follow the method used in \cite{Soda} to analyze the
stability of the internal compacted space. Since internal space is
the direct product of the torus, we could simplify the problem a
little by considering only one of the three tori.  More precisely,
we consider a 6 dimensional space with 2 dimension toroidal
compactified. Let us use the metric ansatz in
(\ref{cosmological_metric}):
\begin{equation}
ds^{2} = g_{\mu\nu}(x^{\mu})dx^{\mu}dx^{\nu}+ds^{2}_{torus} .
\label{simplify_metric}
\end{equation}
The line element for the torus part is written as
\begin{equation}
 ds^2_{torus} = \frac{b^{2}}{\eta}[(dy^{1}+\xi dy^{2})^{2}+\eta^{2} (dy^{2})^{2}],
\label{ansatz}
\end{equation}
with the background antisymmetric field,
\begin{equation}
 B = \left(%
\begin{array}{cc}
  0 & \beta \\
  -\beta & 0 \\
\end{array}%
\right).
 \label{b_ansatz}
\end{equation}
Here we can see that the moduli space is described by 4
parameters, $b^{2}$, $\eta$, $\xi$, and $\beta$.  The parameter
$b$ plays the role of the scale factor of the torus.  While
parameters $\eta$ and $\xi$ control the shape of the moduli.  The
last parameter $\beta$ is called the flux moduli, it determines
the value of the antisymmetric background.

We are interested in heterotic string states whose mass vanishes
at the fixed point. Recall the mass formula for heterotic string
in (\ref{hmass_formula})in $\alpha'=1$ unit:
\begin{equation}
m^{2}=\frac{1}{2}(p_{L}^{2}+p_{R}^{2})+2(N+\tilde{N}-1),
\label{h_mass}
\end{equation}
with matching condition $N-\tilde{N}+1 = m_{A}n^{A} \in
\mathbb{Z}$. $p_{Li}$ and $p_{Ri}$ are defined in
(\ref{m_lattice}). For simplicity, we turn off every oscillators
and zero-modes in SO(32) directions.  This can be done without
affecting the final results.  Since we consider a particular torus
$T^{2}$ subspace of $T^{2}\times T^{2}\times T^{2}$, we must keep
only the vector $p_{Li}$ that is embedded in that subspace.  At
the enhanced symmetry point of toroidal compactification, the
zero-modes $\sqrt{\frac{1}{2}}p_{Li},$ are roots of the simply
laced group $G_{L}$ of rank $2$ with root length $\sqrt{2}$. The
gauge symmetry gets enhanced to $G_L$.

\subsection{$SU(2)\times SU(2)$ fixed point}
By comparing (\ref{ansatz}) and (\ref{b_ansatz}) with
(\ref{G_fixed_point}), it is easy to see that the self-dual fixed
point with enhanced gauge symmetry $G_L = SU(2)\times SU(2)$ has
\begin{equation}
    b^{2}=\eta=1, \beta=\xi=1.
 \label{dual_point}
\end{equation}
We will label the massless string modes by their momenta and
winding numbers in $y^{1}$ and $y^{2}$ directions,
$(m_{1},m_{2},n^{1},n^{2})$.  At the $SU(2)\times SU(2)$
fixed-point, the four heterotic string states that become massless
are:
\begin{center}
\begin{tabular}{ c c c c }

  $(m_{1},$ & $m_{2},$ & $n^{1},$ & $n^{2})$ \\
  \hline
  $\pm 1$ & $0$ & $\pm 1$ & $0$ \\
  $0$ & $\pm 1$ & $0$ & $\pm1$ \\
  \hline
\end{tabular}
\end{center}

At very near to the $SU(2)\times SU(2)$ fixed point, these four
states are not exactly massless and each mode contributes to the
effective potential (\ref{effective_potential}).  In order to
prove that the $SU(2)\times SU(2)$ fixed-point is a stable minimum
of the effective potential, we can consider the minimum of $m^{2}$
in (\ref{effective_potential}).  Let us examine the heterotic
string gas mode by mode.
\\
\\
\textbf{Mode (1, 0, 1, 0)}
\\
This mode corresponds to closed strings wrapped around a circle in
the $y^{1}$ direction. It has the mass squared
\begin{equation}
 m_{1010}^{2} = \frac{1}{b^{2}\eta}(1+\beta\xi)^{2}+\frac{\beta^{2}\eta}{b^{2}} +
 \frac{b^{2}\xi^{2}}{\eta}+\eta b^{2}-2,
 \label{m1010}
\end{equation}
with flat directions $\beta^{2}= \frac{b^{4}\xi^{2}}{\eta^{2}}$
and $1+ \beta\xi=\eta b^{2}$.
\\
\\
\textbf{Mode (0, 1, 0, 1)}
\\
The mass squared for this mode which corresponds to closed strings
wrapped around a circle in the $y^{2}$ direction is given by
%
\begin{equation}
 m_{0101}^{2} = \frac{1}{b^{2}\eta}(\xi-\beta)^{2}+\frac{b^{2}}{\eta} +
 \frac{\eta}{b^{2}}-2.
 \label{m0101}
\end{equation}
There exists the flat directions $\beta = \xi$ and $b^{2}=\eta$.

We find these two flat directions intersect at the self-dual
$SU(2)\times SU(2)$ fixed point $b=\eta=1$, $\xi=\beta=0$. Hence,
by taking into account both modes of string gas, the self-dual
point would be a stable minimum. The stability can be explicitly
verified by expanding the potential around this fixed-point. This
was already discussed by the authors of ~\cite{Soda}. They showed
that, by considering contributions from both massless modes, the
flat directions of the effective potential will disappear if we
move away from the $SU(2)\times SU(2)$ fixed-point.  This implies
that the volume, shape, and flux moduli get stabilized at the self
dual point.  The result will not alter if we add the contributions
from $(-1,0,-1,0)$ and $(0,-1,0,-1)$ modes. Let us investigate
another fixed point.

\subsection{$SU(3)$ fixed point}

Let us consider the fixed point with enhanced symmetry $G_L =
SU(3)$. By comparing (\ref{ansatz}) and (\ref{b_ansatz}) with
(\ref{sample_fixed_point}), this fixed-point has
\begin{equation}
    b^{2}=\eta=\frac{\sqrt{3}}{2}, \beta=\xi=-\frac{1}{2}.
 \label{su3_point}
\end{equation}
At this point on moduli, six heterotic string modes become
massless:
\begin{center}
\begin{tabular}{c c c c}
  $(m_{1},$ & $m_{2},$ & $n^{1},$ & $n^{2})$ \\
  \hline
  $0$ & $\pm 1$ & $0$ & $\pm 1$ \\
  $\pm 1$ & $0$ & $\pm 1$ & $\pm 1$ \\
  $\pm 1$ & $\mp 1$ & $\pm 1$ & $0$ \\
  \hline
\end{tabular}
\end{center}
As we approach the $SU(3)$ fixed-point, these six modes are turned
on.  Their mass will contribute to the effective potential.  Using
the similar investigation to the self-dual fixed-point, it is
enough for us to consider just three massless modes:
\\
\\
\textbf{Mode (0, 1, 0, 1)}
\\
The modes correspond to closed strings wrapped around a circle in
the $y^{2}$ direction are massless at both the $SU(2)\times SU(2)$
and $SU(3)$ fixed-points.  Their contributions to the effective
potential is
\begin{equation}
 m_{0101}^{2} = \frac{1}{b^{2}\eta}(\xi-\beta)^{2}+\frac{b^{2}}{\eta} +
 \frac{\eta}{b^{2}}-2,
 \label{p0101}
\end{equation}
with flat directions $\beta = \xi$ and $b^{2}=\eta$.
\\
\\
\textbf{Mode (1, 0, 1, 1)}
\\
This mode represents string wrapped on both circles of the torus.
Its mass squared can be written as:
%
\begin{equation}
 m_{1011}^{2} = \frac{1}{b^{2}\eta}(1+\beta\xi)^{2}+\frac{\eta\beta^{2}}{b^{2}} + \frac{b^{2}}{\eta}(\xi+1)^{2} +
 \eta b^{2} + \frac{\beta^{2}}{b^{2}\eta}(\beta+2\beta\xi+2)-2.
 \label{p1011}
\end{equation}
It has flat directions $(\xi+1)=-\frac{\beta}{\beta^{2}+b^{4}}$
and $(\xi+1)^{2}=\frac{\eta^{2}\beta^{2}}{b^{4}}$.
\\
\\
\textbf{Mode (1, -1, 1, 0)}
\\
This mode denotes string wrapped on a circle in $y^{1}$. Its mass
squared is:
\begin{equation}
 m_{1-110}^{2} = \frac{1}{b^{2}\eta}(1+\beta\xi)^{2}+\frac{\eta}{b^{2}}(1+\beta)^{2} + \frac{b^{2}}{\eta}(\eta^{2}+\xi^{2}) +
 \frac{\xi^{2}}{b^{2}\eta}(\xi+2\beta\xi+2)-2.
 \label{p1110}
\end{equation}
The mass squared has flat directions
$(\beta+1)=-\frac{\xi}{\xi^{2}+\eta^{4}}$ and
$(\beta+1)^{2}=\frac{b^{4}\xi^{2}}{\eta^{4}}$.

These three flat directions intersect at the $SU(3)$ fixed point:
$b^{2}=\eta=\frac{\sqrt{3}}{2}$ and $\beta=\xi=-\frac{1}{2}$. The
stability can be explicitly verified by perturbing the metric
around the minimum of the effective potential.  For small $m^{2}$,
we get
\begin{equation}
 V = \mu_{4}\sqrt{\frac{g^{ij}p_{i}p_{j}}{g_{s}}}+\frac{1}{2}\frac{\mu_{4}e^{2\phi}}{\sqrt{g^{ij}p_{i}p_{j}}}m^{2}(\beta,b,\eta,\xi).
 \label{potential}
\end{equation}
For convenience, we change coordinates to:
\begin{equation}
b^{2}= \frac{\sqrt{3}}{2}\bar{b}^{2},\,
\xi=-\frac{1}{2}\bar{\xi},\; \eta =
\frac{\sqrt{3}}{2}\bar{\eta},\, \beta=-\frac{1}{2}\bar{\beta}.
\label{bar_coordinates}
\end{equation}
In the new coordinates, the $SU(3)$ fixed-point is at
\begin{equation}
    \bar{b}=\bar{\eta}= \bar{\beta}=\bar{\xi}= 1.
 \label{su3_new_co}
\end{equation}
Let us expand the moduli around the $SU(3)$ fixed-point:
\begin{equation}
\bar{b}= 1+\delta\bar{b},\, \bar{\xi}=1+\delta\bar{\xi},\,
\bar{\eta} = 1+\delta\bar{\eta},\; \bar{\beta}=
1+\delta\bar{\beta}. \label{bar_coords}
\end{equation}
We get
%
\begin{eqnarray}
\delta m_{0101}^{2}&=&
\frac{1}{3}(\delta\bar{\xi}-\delta\bar{\beta})^{2}+(\delta\bar{\eta}-2\delta\bar{b})^{2},\\
\delta m_{1011}^{2}&=& \frac{1}{3}(\delta\bar{\xi}^{2}+
\delta\bar{\xi}\delta\bar{\beta}+\delta\bar{\beta}^{2})+
(\delta\bar{\eta}^{2}+2\delta\bar{\eta}\delta\bar{b}+4\delta\bar{b}^{2}) + \delta\bar{\eta}\delta\bar{\beta}- 2\delta\bar{\xi}\delta\bar{b},\\
\delta m_{1-110}^{2}&=& \frac{1}{3}(\delta\bar{\xi}^{2}+
\delta\bar{\xi}\delta\bar{\beta}+\delta\bar{\beta}^{2})+
(\delta\bar{\eta}^{2}+2\delta\bar{\eta}\delta\bar{b}+4\delta\bar{b}^{2})
- \delta\bar{\eta}\delta\bar{\beta}+
2\delta\bar{\xi}\delta\bar{b}. \label{bar_coords}
\end{eqnarray}
We can easily observe that each potential has flat directions.
However, by summing all modes together, flat directions disappear
in the total potential:
%
\begin{equation}
\delta m_{su(3)}^{2}= \delta m_{0101}^{2}+\delta
m_{1011}^{2}+\delta m_{1-110}^{2} =
\delta\bar{\beta}^{2}+\delta\bar{\xi}^{2}+3\delta\bar{\eta}^{2}+3\delta\bar{b}^{2}.
\label{totalv3}
\end{equation}
This implies that the $SU(3)$ fixed point is a stable fixed point
as we expected.

Let us summarize our results.  The moduli can be stabilized at
both fixed-points depending on the initial conditions.  At the
$SU(2) \times SU(2)$ fixed-point four modes, $(\pm1,0,\pm1,0)$ and
$(0,\pm1,0,\pm1)$, are massless. If we move away from the fixed
point, these modes become massive and contribute to the
contracting potential.  On the other hands, if the initial
condition is near to the $SU(3)$ fixed point, we should turn off
the $(0,\pm1,0,\pm1)$ modes as they are massive at the $SU(3)$
fixed point and will not contribute to the stabilization
mechanism. The six massive modes $(\pm1,0,\pm1,0)$,
$(\pm1,0,\pm1,\pm1,)$ and $(\pm1,\mp1,\pm1,0)$ will be turned on
and play their roles in a contracting potential to stabilize the
moduli at the $SU(3)$ fixed point.

We also have to consider the stability of the dilaton field.  This
can be done by considering the equations of motion from variations
of the string frame action in (\ref{eff_action_1}).  For
simplicity, we will fix the shape of the torus and give
constraints to the flux moduli as:

\begin{equation}
\frac{b^{2}}{\eta}=e^{2\upsilon},\;\beta=\xi\frac{b^{2}}{\eta}=
\frac{z}{2}\:e^{2\upsilon},\;\xi^{2}+\eta^{2}=1.
\label{fix_moduli}
\end{equation}
Here $\upsilon$ is the function of the comoving time $t = x^{0}$
and $z$ is a constant parameter with $z = 0$ and $z = -1$
representing $SU(2)\times SU(2)$ and $SU(3)$ fixed-points
respectively.  The 6 dimensional metric in (\ref{simplify_metric})
can be written as:
\begin{equation}
ds^2 = -dt^{2}+e^{2\lambda(t)}\sum_{i=1}^{4}dx^{i}dx^{i} +
e^{2\upsilon(t)}[(dy^{1})^{2}+ z dy^{1}dy^{2} + (dy^{2})^{2}],
\label{ansatz2}
\end{equation}
with the background antisymmetric field,
\begin{equation}
 B(t) = \frac{e^{2\upsilon(t)}}{2}\left(%
\begin{array}{cc}
  0 & z \\
  -z & 0 \\
\end{array}%
\right).
 \label{b_ansatz2}
\end{equation}
The equations of motion are:
%
\begin{eqnarray}
\ddot{\lambda} -
2\dot{\lambda}\dot{\phi}+3\dot{\lambda}^{2}+2\dot{\lambda}\dot{\upsilon}&=&\kappa^{2}P_{\lambda}
e^{2\phi}\\
\ddot{\upsilon}-2\dot{\upsilon}\dot{\phi}+3\dot{\lambda}\dot{\upsilon}+\frac{8}{4-z^{2}}\:\dot{\upsilon}^{2}&=&\kappa^{2}P_{\upsilon}e^{2\phi}\\
2\ddot{\phi}-4\dot{\phi}^{2}+6\dot{\phi}\dot{\lambda}+4\dot{\phi}\dot{\upsilon}+\frac{z^{2}}{4-z^{2}}\dot{\upsilon}^{2}&=&\frac{\kappa^{2}}{2}T\:e^{2\phi}
 , \label{eom1}
\end{eqnarray}
where $T = 3P_{\lambda}+2P_{\upsilon}-\rho$ is the trace of the
energy momentum tensor of string gas, $P_{\lambda}(P_{\upsilon})$
denotes pressure in non-compact (compact) directions ,and a dot
denotes a derivative with respect to time. By writing equation
(\ref{eom1}) as:
\begin{equation}
\frac{d^{2}}{dt^{2}}(e^{-2\phi})+3\dot{\lambda}\frac{d}{dt}(e^{-2\phi})+2\dot{\upsilon}\frac{d}{dt}(e^{-2\phi})+\frac{1}{2}\frac{z^{2}}{4-z^{2}}\dot{\upsilon}^{2}(e^{-2\phi})=-\kappa^{2}T,
\label{eom2}
\end{equation}
We come to the same conclusion as \cite{Soda}.  At the fixed-point
$\upsilon=0$ and $z=0,-1$, the energy-momentum tensor of the
string gas is traceless $T = 0$. The dilaton is stabilized by the
hubble damping term due to the expansion of the non-compact
direction.  Since the dilaton can move along the flat direction,
it is marginally stable at both fixed points.


\section{Conclusions}
\label{sec:conclusions}

We have shown that, for gas of string compactified on torus with
an antisymmetric background field, there exist other fixed-points
apart from the self-dual fixed-point.  We have examined the toy
model of heterotic string gas compactified on the 6 dimensional
internal space.  We assume the compact space to be the direct
product of three 2 dimensional tori.  The stability of each torus
can be investigated separately.

In the moduli of 2-dimensional torus, there are two fixed points.
The first is the self-dual fixed point of T-duality.  At this
fixed-point, the heterotic string spectrum has  $SU(2) \times
SU(2)$ gauge symmetry.  The second fixed point produces $SU(3)$
enhanced gauge symmetry to the string spectrum and it is a fixed
point for the more general target space duality.  More precisely,
it is the fixed point for T-duality modulo the discrete rotational
group and a shift on the $B$ field with a constant integer value.
We have shown that these two points are stable fixed-points on the
moduli.  The dilaton is marginally stable at both fixed-points.
The existence of these fixed-points might have some application to
cosmology.  We leave this for future investigations.

\section*{Acknowledgements}

The author would like to thank Prof. A.C. Davis and Dr. A.
Ungkitchanukit for discussions and for kindly reading the
manuscript. We are grateful to Prof. K.S. Narain for useful
discussions. This work is partly supported by Chulalongkorn
University and ICTP through the Junior Associate fellowship.


\begin{thebibliography}{99}
 \bibitem{BV}
     R.H.~Brandenberger and C.~Vafa, Nucl.Phys. \textbf{B316} (1989) 391.
 \bibitem{T-dual}
     K.~Kikkawa and M.~Yamasaki, Phys.Lett. \textbf{B149} (1984) 357.
 \bibitem{TV}
     A.A.~Tseytlin and C.~Vafa, Nucl.Phys. \textbf{B372} (1992) 443 [arXiv:hep-th/9109048].
 \bibitem{KP}
     J.~Kripfganz and H.~Perlt, Class. Quant. Grav. \textbf{5} (1988) 453.
 \bibitem{Greene1}
     R.~Easther, B.R.~Greene and M.G.~Jackson, Phys.Rev. \textbf{D66} (2002) 023502 [arXiv:hep-th/0204099 ]
 \bibitem{alex}
     S. Alexander, R.H.~Brandenberger and D.~Easson, Phys.Rev. \textbf{D62} (2000) 103509 [arXiv:hep-th/0005212].
 \bibitem{WB1}
     S.~Watson and R.H.~Brandenberger, Phys.Rev. \textbf{D67} (2003) 043510 [arXiv:hep-th/0207168].
 \bibitem{Boehm}
     T.~Boehm and R.~Brandenberger, JCAP \textbf{0306}, (2003) 008 [arXiv:hep-th/0208188].
 \bibitem{Easson}
     E.A.~Easson, Int. J. Mod. Phys. \textbf{A18} (2003) 4295 [arXiv:hep-th/0110225].
 \bibitem{Bassett}
     B.A.~Bassett, M.~Borunda, M.~Serone and S.~Tsujikawa, Phys.Rev. \textbf{D67} (2003) 123506 [arXiv:hep-th/0301180].
 \bibitem{Campos}
     A. Campos, Phys. Rev. \textbf{D68}, (2003) 104017 [arXiv:hep-th/0304216].
 \bibitem{Greene2}
     R.~Easther, B.R.~Greene, M.G.~Jackson and D.~Kabat, JCAP \textbf{0502} (2005)
      009 [arXiv:hep-th/0409121]
 \bibitem{Trodden}
     E.A.~Easson and M.~Trodden, Phys.Rev. \textbf{D72} (2005) 026002 [arXiv:hep-th/0505098].
 \bibitem{Kaya}
     A.~Kaya, Phys.Rev. \textbf{D72} (2005) 066006 [arXiv:hep-th/0504208].
 \bibitem{moduli}
     R.H.~Brandenberger, \emph{Moduli Stabilization in String Gas Cosmology}, arXiv:hep-th/0509159.
 \bibitem{WB2}
     S.~Watson and R.H.~Brandenberger, JCAP \textbf{0311} (2003) 008.[arXiv:hep-th/0307044]
 \bibitem{Battefeld}
     T.~Battefeld and S.~Watson, JCAP \textbf{0406} (2004) 001 [arXiv:hep-th/0403075].
 \bibitem{Patil}
     S.P.~Patil and R.H.~Brandenberger, Phys.Rev. \textbf{D71} (2005) 103522 [arXiv:hep-th/0401037].
 \bibitem{Watson}
     S.~Watson, Phys.Rev. \textbf{D70} (2004) 066005 [arXiv:hep-th/0404177].
 \bibitem{Soda}
     S. Kanno and J. Soda, Phys.Rev. \textbf{D72} (2005) 104023 [arXiv:hep-th/0509074].
 \bibitem{Narain}
     K.S.~Narain, Phys.Let. \textbf{B169} (1986) 41.
 \bibitem{NSW}
     K.S.~Narain, M.H.~Sarmadi and E.~Witten, Nucl.Phys. \textbf{B279} (1987) 369.
 \bibitem{GPR}
     A.~Giveon, M.~Porrati and E.~Rabinovici, Phys.Rept. \textbf{244} (1994) 77 [arXiv:hep-th/9401139].
 \bibitem{EGRS}
     S.~Elitzur, E.~Gross, E.~Rabinovici and N.~Seiberg, Nucl.Phys. \textbf{B283} (1987) 413.
 \bibitem{GRV}
     A.~Giveon, E.~Rabinovici and G.~Veneziano, Nucl.Phys. \textbf{B322} (1989) 167.
 \bibitem{SW}
     A.~Shapere and F.~Wilczek, Nucl.Phys. \textbf{B320} (1989) 669.
 \bibitem{GV}
     M.~Gasperini and G.~Veneziano, Phys.Lett. \textbf{B277} (1992) 256.
 \bibitem{auttakit}
     A.~Chattaraputi, F.~Englert, L.~Houart and A.~Taormina, JHEP \textbf{0209} (2002) 037.
\bibitem{Witten}
     E.~Witten, Nucl.Phys. \textbf{B195} (1982) 481.

\end{thebibliography}
\end{document}